\documentclass{elsarticle}

\usepackage{amsmath}
\usepackage{amsfonts}
\usepackage{amssymb}
\usepackage{graphicx}

\begin{document}

\begin{frontmatter}

\title{ Does nuclear matter bind at large $N_{c}$?}
\author{Luca Bonanno and Francesco Giacosa}

\address{Institut f\"{u}r Theoretische Physik \\
Johann Wolfgang Goethe - Universit\"{a}t\\ 
Max von Laue--Str. 1 D-60438 Frankfurt, Germany}

\begin{abstract}
The existence of nuclear matter at large $N_{c}$ is investigated in the
framework of effective hadronic models of the Walecka type. This issue  is
strongly related to the nucleon-nucleon attraction in the scalar channel, and
thus to the nature of the light scalar mesons. Different scenarios for the
light scalar sector correspond to different large $N_{c}$ scaling properties
of the parameters of the hadronic models. In all realistic phenomenological
scenarios for the light scalar field(s) responsible for the attraction in the
scalar channel it is found that nuclear matter does not bind in the large
$N_{c}$ world. We thus conclude that $N_{c}=3$ is in this respect special: $3$
is fortunately not large at all and allows for nuclear matter, while large
$N_{c}$ would not.
\end{abstract}

\end{frontmatter}

\section{Introduction}

The large $N_{c}$ limit constitutes a well-defined and systematic
theoretical framework to address fundamental questions of QCD \cite%
{thooft,witten,reviewnc}. Recently, basic properties of the QCD phase
diagram for $N_{c}\gg1$ have been presented in Ref.~\cite{quarkyonic} and
further investigated in Refs.~\cite{sasaki,torrieri,dicoto,kojo} and refs.
therein.

In this work we address the following question: does nuclear matter bind for 
$N_{c}\gg1?$ It is interesting to investigate whether the existence of
nuclear matter is a special phenomenon of $N_{c}=3$ or it is a general
feature independent on the number of colors.

It is not possible to answer this question by using QCD. In fact, the latter
is not solvable, not even in the large $N_{c}$ limit. Thus, the only way to
study nuclear matter at large $N_{c}$ is to use an effective Lagrangian. The
most convenient choice for this purpose is the Walecka model \cite{walecka},
which has been widely used for studies at nonzero density. In fact, although
this model is naive, we use it for several reasons: (i) Regardless of the
real model describing nature, nuclear matter saturation can be always
described by using an effective Walecka-like model including attractive
scalar interactions and repulsive vector interactions. This means that the
(unfortunately unknown) correct chiral model for low-energy hadrons must
reduce, for densities close to nuclear matter and for small temperature, to
a Walecka-like model. (ii) Although the Lagrangian does not embed important
symmetries of QCD such as chiral symmetry and scale invariance, we are
limiting our study to nuclear matter. This is a regime of baryon densities
at which these symmetries are strongly broken. (iii) The Walecka model has
the advantage of being simple, allowing a direct understanding of the large $%
N_{c}$ behavior of nuclear matter.

For $N_{c}=3$ the couplings are fixed to recover the usual nuclear density
properties: saturation at $\rho_{0}=0.16$ fm$^{-3}$ and an energy per
nucleon $E/A=-16$ MeV. We then rescale the couplings with appropriate powers
of $N_{c}$ and perform the study at higher $N_{c}$ values: in this way it is
possible to check if nuclear matter still exists when increasing $N_{c}.$

We shall find that the binding of nuclear matter at large $N_{c}$ strongly
depends on the nature of the lightest scalar resonance(s) \cite%
{amslerrev,dynrec}. In the (old-fashioned) assignment in which the lightest
scalar resonance $f_{0}(600)$ is predominantly a quark-antiquark state,
nuclear matter binds indeed at each $N_{c}$. Moreover, the binding energy
increases for increasing $N_{c}$.

However, the quark-antiquark scenario is regarded as unfavored by most
recent works on the low-energy scalar sector \cite%
{dynrec,pelaeznc,pennington,denis}. We thus study alternative assignments
for the scalar resonances, which are in agreement with the phenomenology.
For instance, the so-called tetraquark scenario can describe well the
properties of the light scalar states below $1$ GeV, see the original work
of Ref.~\cite{jaffeorig} and further investigations in Refs.~\cite%
{exotica,varietq,fariborz,tqmio}. Interpreting $f_0(600)$ as a tetraquark
state, and taking into account the correct large $N_{c}$ limit of these
objects (Sec. 2.4), it is found that nuclear matter does \emph{not} bind for 
$N_{c}\gg3,$ and indeed ceases to exist already for $N_{c}=4.$

We also repeat our analysis for other assignments in the scalar sector: (i)
Even if the light scalar states are predominantly tetraquarks, scalar
quarkonium states must exist. For this reason we study an enlarged scenario
in which, besides the resonance $f_{0}(600)$ interpreted as predominantly
tetraquark, a second scalar field, identified with the resonance $%
f_{0}(1370) $, is added and interpreted as predominantly quarkonium. This
scenario is in agreement with the outcome of Ref. \cite{ml}, where two
scalar fields with similar masses are needed in order to describe
nucleon-nucleon scattering data. (ii) The role of two-pion-exchange (TPE)
processes can be important for nuclear matter \cite{weisenucl}. We thus test
the scenario in which a scalar field describes effectively the TPE
attraction in the scalar-isoscalar channel. (iii) The assignment in which
the lightest scalar state $f_{0}(600)$ is interpreted as a glueball is
investigated. (iv) The lightest scalar resonance $f_{0}(600)$ can be also
regarded as a `dynamically generated resonance' emerging from pion-pion
interactions \cite{pelaeznc}.

It is remarkable that also in all these cases nuclear matter does \emph{not}
bind for in the large $N_{c}$ limit. The non existence of nuclear matter for
large $N_{c}$ is thus not an artifact of a particular assignment for the
light scalar states, but is a stable result as soon as the quarkonium
interpretation is abandoned. Note also that, while studies of
nucleon-nucleon potentials in the large $N_{c}$ limit exist \cite%
{manohar,arriola}, the different scaling properties for $N_{c}\gg 3$ due to
the non-quarkonium nature of the scalar attraction have not yet been (to our
knowledge) systemically investigated.

The paper is organized as follows: in Sec.~2 we investigate the existence of
nuclear matter for $N_{c}\gg1$ for different scenarios for scalar resonances
and in Sec.~3 we draw our conclusions.

\section{Nuclear matter at large $N_{c}$}

\subsection{The Walecka model}

The Lagrangian of the Walecka model reads \cite{walecka}: 
\begin{eqnarray}
\mathcal{L}&=&\bar{\psi}[\gamma^{\mu}(i\partial_{\mu}-g_{\omega}\omega_{\mu
})-(m_{N}-g_{\sigma}\sigma)]\psi+\frac{1}{2}\partial^{\mu}\sigma\partial_{%
\mu }\sigma-\frac{1}{2}m_{\sigma}^{2}\sigma^{2}-\frac{1}{4}%
F^{\mu\nu}F_{\mu\nu }  \notag \\
&+&\frac{1}{2}m_{\omega}^{2}\omega^{\mu}\omega_{\mu}-V_{\sigma}(\sigma)\text{
,}  \label{lagrdiq}
\end{eqnarray}
where $\psi$ and $\omega_{\mu}$ are the nucleon and the isoscalar-vector
field, respectively, and $F_{\mu\nu}=\partial_{\mu}\omega_{\nu}-\partial_{%
\nu }\omega_{\mu}$. Finally, $\sigma$ is a scalar field, usually identified
with the lightest scalar resonance: $f_{0}(600).$ $V_{\sigma}(\sigma)$ is a
potential containing self-interaction terms of the $\sigma$ field, which we
do not consider here for simplicity.

The mean values of the scalar and vector condensates can be easily found by
solving the Euler-Lagrange equations: 
\begin{equation}
\bar{\sigma}=\left( \frac{g_{\sigma}}{m_{\sigma}}\right) ^{2}\rho_{s}\text{
, }\bar{\omega}_{0}=\left( \frac{g_{\omega}}{m_{\omega}}\right) ^{2}\rho _{B}%
\text{ ,}
\end{equation}
where $\rho_{s}$ and $\rho_{B}$ are the scalar density and the baryon
density, respectively.

\subsection{The large $N_{c}$ limit}

We briefly summarize basic properties of hadrons in the large $N_{c}$ limit.
These features have been originally investigated in Refs.~\cite%
{thooft,witten} and reviewed in Ref.~\cite{reviewnc}. The following
properties at large $N_{c}$ have been deduced:

\begin{itemize}
\item Quark-antiquark meson masses have a smooth limit for large $N_{c}$: $%
m_{\bar{q}q}\propto N_{c}^{0}.$ Moreover, the general $n$-point interaction
vertex is of order $N_{c}^{-(n-2)/2}$, therefore the quark-antiquark states
are non-interacting particles when $N_{c}\rightarrow\infty$ .

\item The baryon mass is of order $N_{c}$.

\item The baryon-meson amplitude is of order $\sqrt{N_{c}}$ and the
baryon-baryon interaction, mediated by meson exchange, is of order $N_{c}.$

\item Four-quark states do not survive in the large $N_{c}$ limit, but a
different limit for tetraquark states can be considered, see Sec. 2.4 for
details.
\end{itemize}

\subsection{The naive quark-antiquark scenario}

We start by considering the case in which the medium-range attraction among
nucleons is only mediated by the exchange of a quark-antiquark state, that
can be identified with the lightest scalar resonance $f_{0}(600).$ This is
the old-fashioned assignment for low-energy effective models of QCD, such as
the linear $\sigma$ model \cite{lee,geffen}, and the Nambu Jona-Lasinio
model \cite{njl}.

It should be noticed already at this stage that this assignment is unfavored
by most recent studies of light scalar mesons \cite%
{dynrec,pelaeznc,pennington}. Also in the updated version of the linear $%
\sigma$ model with (axial)vector mesons of Ref.~\cite{denis}, it is found
that the quark-antiquark scalar state $\sqrt{\frac{1}{2}}(\bar{u}u+\bar{d}d)$
corresponds to the resonance $f_{0}(1370),$ rather than to the resonance $%
f_{0}(600)$. Nevertheless, due to its historical importance and the
well-defined large $N_{c}$ behavior, we first investigate this scenario.

\begin{figure}[h]
\begin{centering}
\includegraphics*[scale=0.4]{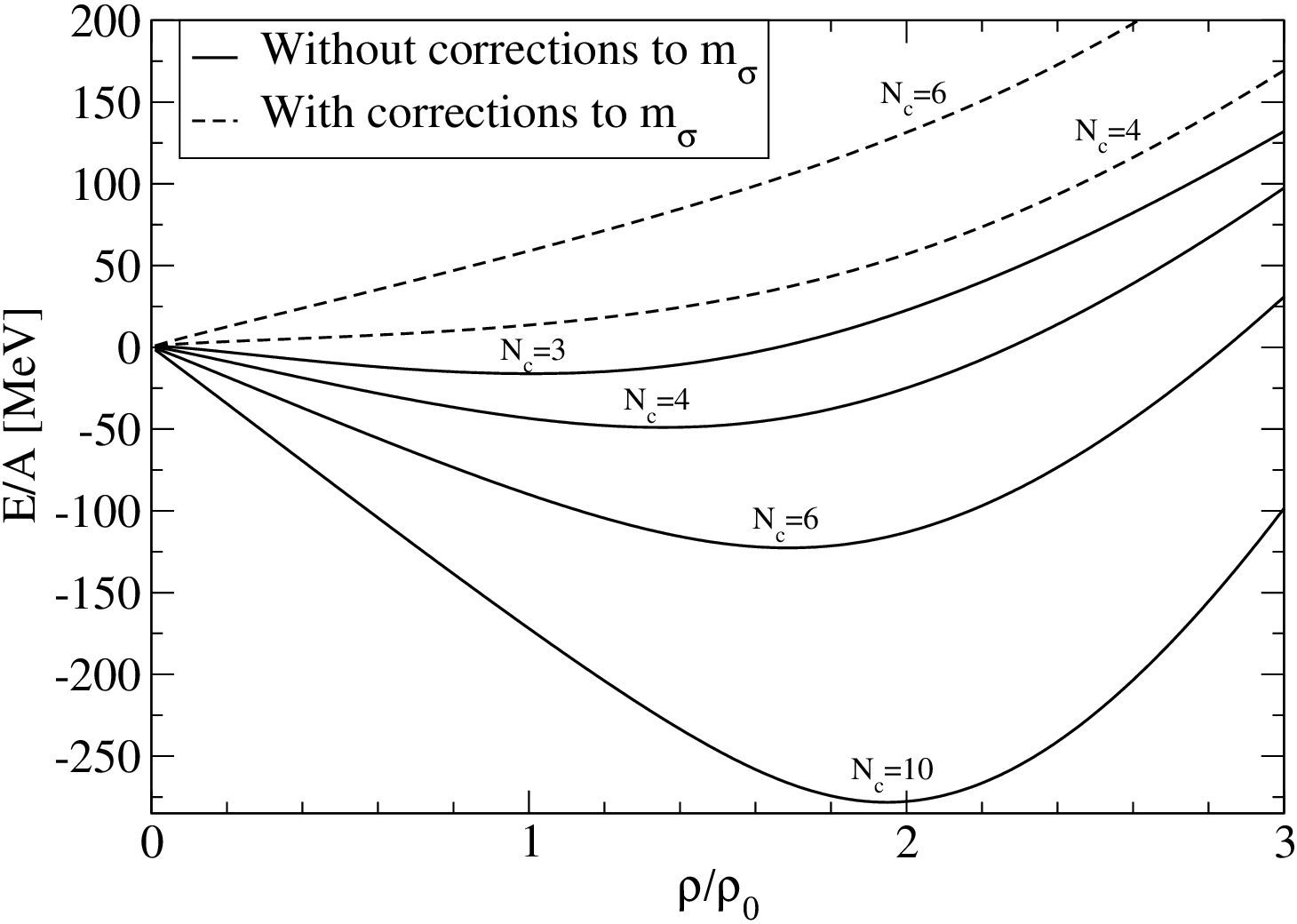}\caption{Binding energy per
nucleon as a function of the baryon density (in units of $\rho_{0}=0.16$fm$^{-3}$)
for different values of $N_{c}$. The solid lines refer to the
case in which no loop corrections to $m_{\sigma}$ are considered; the dashed
lines result from the inclusion of these corrections, by using
Eq.~(\ref{msigmancnew}) with $b_{\sigma}=953$ MeV.}%
\label{be}%
\end{centering}
\end{figure}

\begin{figure}[h]
\begin{centering}
\includegraphics*[scale=0.4]{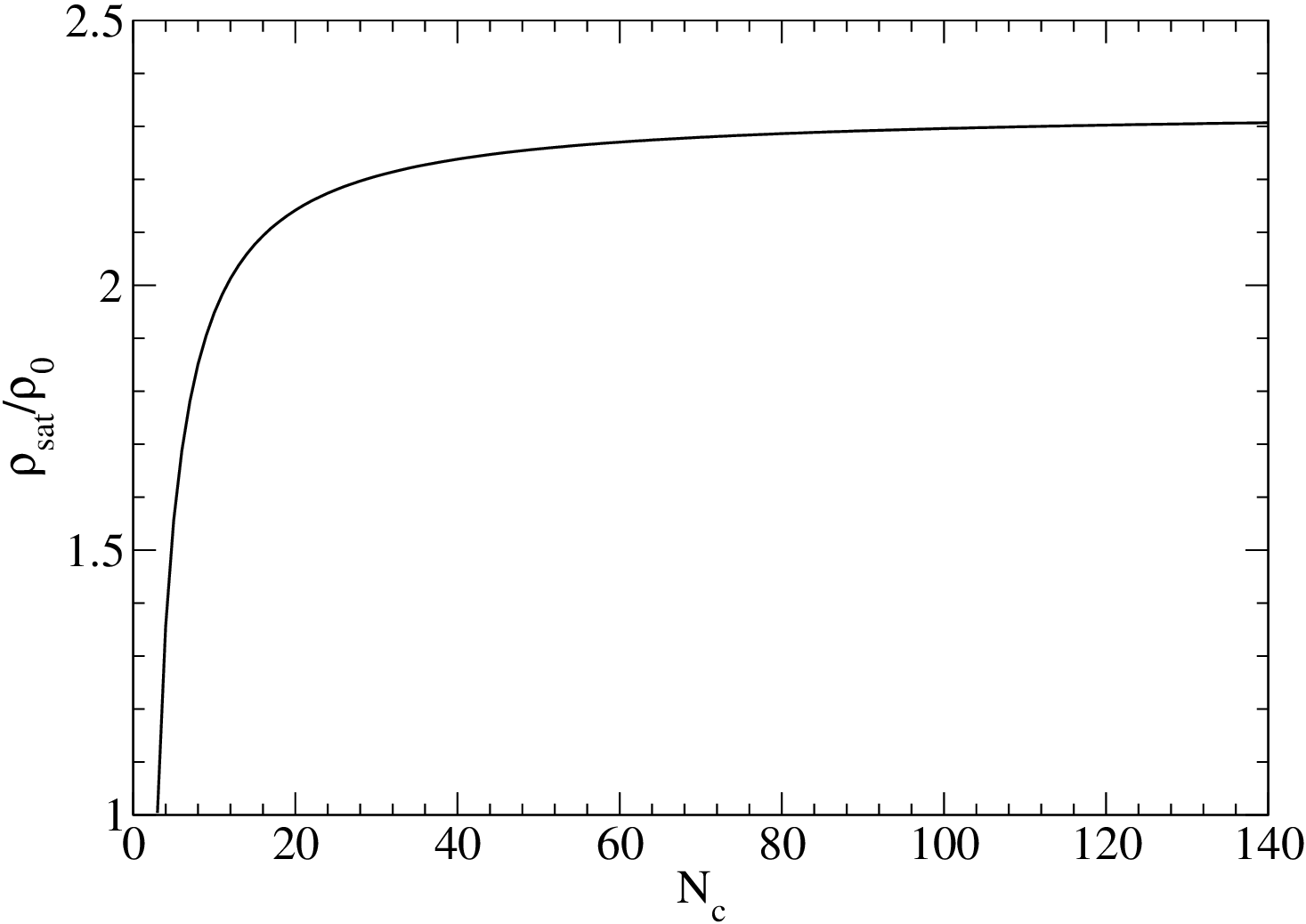}\caption{Baryon density at
saturation (in units of $\rho_{0}=0.16$fm$^{-3}$) as a function of the number of
colors.}%
\label{rho}%
\end{centering}
\end{figure}

\begin{figure}[h]
\begin{centering}
\includegraphics*[scale=0.4]{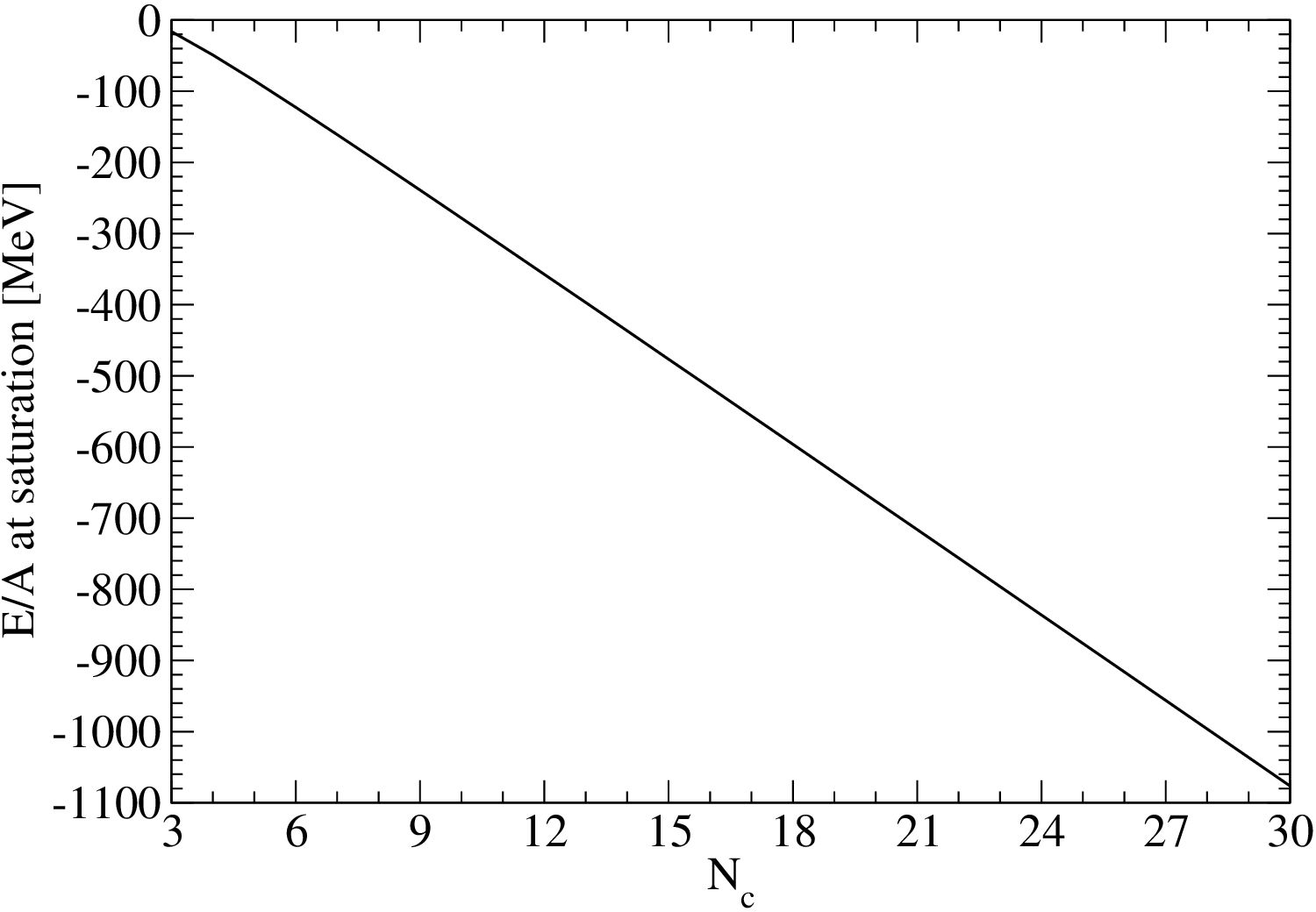} .\caption{Binding energy per
nucleon at saturation as a function of the number of colors.}%
\label{en}%
\end{centering}
\end{figure}

Following the scaling properties in Sec. 2.2, we can easily deduce the
following scaling laws for the masses and the couplings of the model in Eq. (%
\ref{lagrdiq}): 
\begin{align}
m_{\sigma} & \longrightarrow m_{\sigma}\text{ ;}  \label{msigmanc} \\
m_{\omega} & \longrightarrow m_{\omega}\text{ , }m_{N}\longrightarrow m_{N}%
\frac{N_{c}}{3}\text{ ;}  \label{omega} \\
g_{\sigma} & \longrightarrow g_{\sigma}\sqrt{\frac{N_{c}}{3}}\text{ , }
g_{\omega}\longrightarrow g_{\omega}\sqrt{\frac{N_{c}}{3}}\text{ .}
\label{gis}
\end{align}
In order to understand if nuclear matter still exists for $N_{c}>3$, we
computed the equation of state (EoS) of cold, isospin symmetric nuclear
matter, making use of the scaling relations Eqs.~(\ref{msigmanc}), (\ref%
{omega}) and (\ref{gis}). In order to reproduce the saturation, we use the
following numerical values of the parameters at $N_{c}=3$: $%
g_{\sigma}^{2}/4\pi=11.10$ with $m_{\sigma}=600$ MeV and $%
g_{\omega}^{2}/4\pi=14.37$ with $m_{\omega}=783$ MeV. The results are shown
in Fig.~\ref{be}, where it is clear that nuclear matter reaches saturation
for all values of $N_{c}$. The reason for this property is easy to
understand: since in this case $\omega_{\mu}$ and $\sigma$ are both $q\bar{q}
$ states, their couplings scale in the same way. Therefore, the balance
between attraction and repulsion among nucleons, leading to saturation, does
not depend on the number of colors. As a consequence of the scaling, the
value of the saturation density, as shown in Fig.~\ref{rho}, reaches an
asymptotic value of $\sim2.3\rho_{0}$ when $N_{c}\rightarrow\infty$. On the
other hand, since the energy of the system is of order $N_{c}$, the binding
energy per nucleon at saturation must grow linearly with $N_{c}$. This
behavior is shown in Fig.~\ref{en}.

\begin{figure}[h]
\begin{centering}
\includegraphics*[scale=0.4]{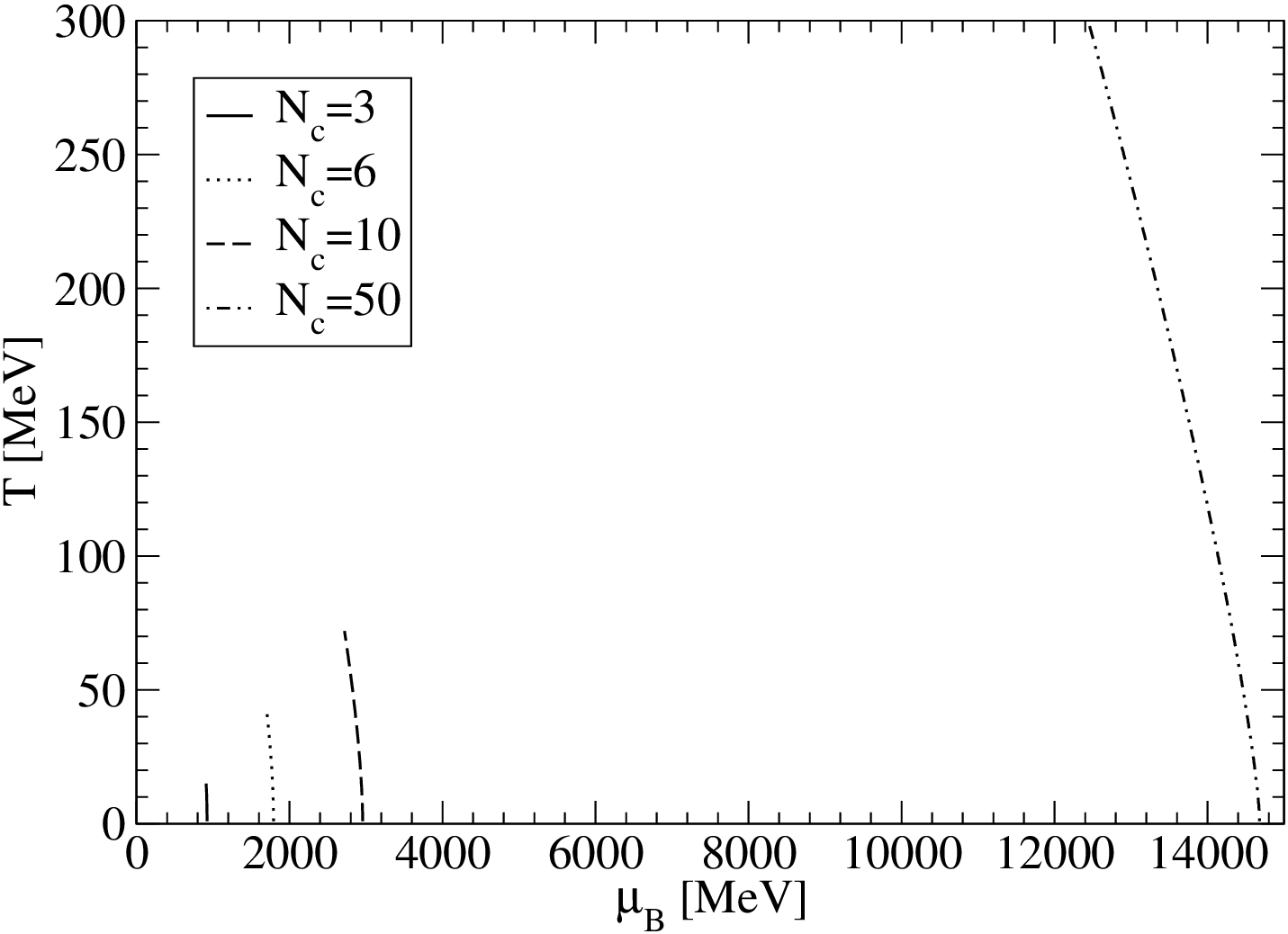}\caption{Liquid-gas transition critical lines for different values
of the number of colors ($N_c=3, 6, 10, 50$).}%
\label{gas}%
\end{centering}
\end{figure}

In Fig. \ref{gas} we show the liquid-gas critical lines for different values
of $N_{c}.$ Not only the critical baryon chemical potential, but also the
critical temperature grows linearly with $N_{c}$. We reach a domain of
temperatures in which our present mean-field and Walecka-based study should
be regarded with care. Moreover, for $N_{c}$ large enough, the critical
temperature would also overshoot the deconfinement temperature $%
T_{dec}\sim\Lambda_{QCD},$ which is a $N_{c}$-independent quantity. It is
anyhow interesting to observe that in this scenario the nuclear-liquid
transition line becomes longer on the $T$-direction, see also the discussion
in Ref.~\cite{torrieri}, in which a Van Der Waals approach to describe the
nuclear-liquid phase transition is used.

The results presented in this subsection lead -- at first sight -- to a
strongly bound nuclear matter for large $N_{c}.$ However, even in the
quark-antiquark scenario for the light $\sigma$ meson, the next-to-leading
order corrections to the $\sigma$ mass are non-negligible. Namely, the
function $m_{\sigma}^{2}(N_{c})$ can be rewritten as%
\begin{equation}
m_{\sigma}^{2}(N_{c})=m_{\sigma}^{2}+b_{\sigma}^{2}\left( \frac{1}{3}-\frac{1%
}{N_{c}}\right) \text{ .}  \label{msigmancnew}
\end{equation}
This formula takes into account the fact that the mass for $N_{c}=3,$ $%
m_{\sigma}^{2}(N_{c}=3)=m_{\sigma}^{2},$ is reduced by meson loops with
respect to the large $N_{c}$ asymptotic value $m_{\sigma}^{2}(N_{c}\gg1)=m_{%
\sigma}^{2}+b_{\sigma}^{2}/3$ \cite{lupo}. In the case of the light $%
f_{0}(600)$ meson this mass reduction is generated by the pion loops and is
large, as the large width of this resonance confirms. Numerically, one has $%
b_{\sigma}^{2}\simeq3(350$-$600$ MeV$)^{2},$ which implies a meson-loop mass
reduction of about 100-250 MeV.

Note that the same modification holds in principle also for the $\omega$
meson, but it is negligibly small due to the very small width of this
resonance. This modification can be safely omitted.

When using the more realistic Eq.~(\ref{msigmancnew}) instead of Eq.~(\ref%
{msigmanc}) one already observes that the nuclear matter does not bind at $%
N_{c}>3$. This property is shown in Fig.~\ref{be}, where the dashed lines
represent the binding energy per nucleon when Eq.~(\ref{msigmancnew}) is
used.

\subsection{The tetraquark scenario}

The tetraquark scenario offers a consistent scheme to interpret the scalar
states below $1$ GeV \cite{jaffeorig,exotica,varietq,fariborz,tqmio}. The
state $f_{0}(600)$ is described as a bound state of two `good' diquarks: $%
f_{0}(600)=\frac{1}{2}[\bar{u},\bar{d}][u,d],$ where the commutation means
anti-symmetrization in flavor space. Similarly, the color wave function of
the tetraquark is given by%
\begin{equation}
\lbrack\bar{R},\bar{B}][R,B]+[\bar{G},\bar{B}][G,B]+[\bar{R},\bar{G}][R,G]%
\text{ ,}
\end{equation}
where $R,B,G$ refer to the three colors of the quark.

The term `good' diquark \cite{exotica} signalizes the flavor and color
antisymmetric spinless diquark, in which a particularly strong attraction
takes place \cite{variedq}. The very same good diquark is also expected to
be an important piece of the nucleon described as a quark-diquark bound
state. It is also a basic object for color superconductivity at large
density.

The relevant question for the present study is the description of a
tetraquark in the large $N_{c}$ limit. It is actually known that a
tetraquark, which is made of two quarks and two antiquarks, does not survive
in the large $N_{c}$ limit. This fact was already recognized in the original
works by 't Hooft \cite{thooft} and Witten \cite{witten}: instead of a
tetraquark one has in the large $N_{c}$ limit two `standard' quark-antiquark
mesons.

There is however another object which can be considered in the Large $N_{c}$
limit. Considering that our tetraquark for $N_{c}=3$ consists of two good
diquarks, it should be asked what is a good diquark for $N_{c}>3$. The
answer is simple: a `good diquark' for $N_{c}>3$ is an object with $N_{c}-1$
quarks in a antisymmetric color wave function: 
\begin{equation}
d_{a_{1}}=%
\varepsilon_{a_{1}a_{2}a_{3}...a_{N_{c}}}q^{a_{2}}q^{a_{3}}...q^{a_{N_{c}}}%
\text{ with }a_{2},...a_{N_{c}}=1,...N_{c}\text{ .}
\end{equation}
This means that the generalization of a `tetraquark' for $N_{c}>3$ is a
bound state of a $(N_{c}-1)$-quark object and a $(N_{c}-1)$-antiquark object 
\cite{dynrec,liu}: 
\begin{equation}
\chi=\sum_{a_{1}=1}^{N_{c}}d_{a_{1}}^{\dagger}d_{a_{1}}\text{ .}
\end{equation}
In this way also the `tetraquark' has a well-defined limit for $N_{c}\gg1,$
whose mass scales as 
\begin{equation}
m_{\chi}\varpropto2(N_{c}-1)\sim N_{c}\text{ .}
\end{equation}
The interaction of the generalized tetraquark $\chi$ with mesons and with
nucleons is different:

(i) The formation of two quark-antiquark mesons requires the annihilations
of $(N_{c}-3)$ quark-antiquark pairs. For this reason the amplitude for the
process $\chi\rightarrow\bar{Q}Q,$ where $Q$ represents a quark-antiquark
state, is given by $A_{\chi\rightarrow\bar{Q}Q}$ $\propto p^{N_{c}-3}\sim
p^{N_{c}}\sim e^{-N_{c}},$ where $p$ is the annihilation probability of a
quark from $d_{a_{1}}$ and an antiquark from $d_{a_{1}}^{\dagger}.$ The full
decay width of the process $\chi\rightarrow\bar{Q}Q$ as a function of $N_{c}$
reads: 
\begin{equation}
\Gamma_{\chi\rightarrow\bar{Q}Q}\sim\frac{\sqrt{\frac{m_{\chi}^{2}}{4}%
-m_{Q}^{2}}}{m_{\chi}^{2}}[A_{\chi\rightarrow\bar{Q}Q}]^{2}\sim\frac{1}{N_{c}%
}e^{-N_{c}}\text{ .}  \label{tqmesons}
\end{equation}
It is interesting to observe that $\Gamma_{\chi\rightarrow\bar{Q}Q}$ first
increases as function of $N_{c}$ in virtue of the increasing phase space,
then it starts to decrease exponentially because of the suppressed decay
amplitude. The interaction of this generalized tetraquark with
quark-antiquark mesons assures that this object is well defined in the large 
$N_{c}$ limit, but we will not need the scaling of Eq. (\ref{tqmesons}).

(ii) The decay of the state $\chi$ into two baryons takes place upon
creation of a single quark-antiquark pair from the vacuum. Thus, the
amplitude of this process is proportional to $p,$ i.e.~to $N_{c}^{0}$. The
coupling constant $g_{\chi}$ for the $\chi$-nucleons interaction scales as $%
N_{c}^{0}.$

\begin{figure}[h]
\begin{centering}
\includegraphics*[scale=0.4]{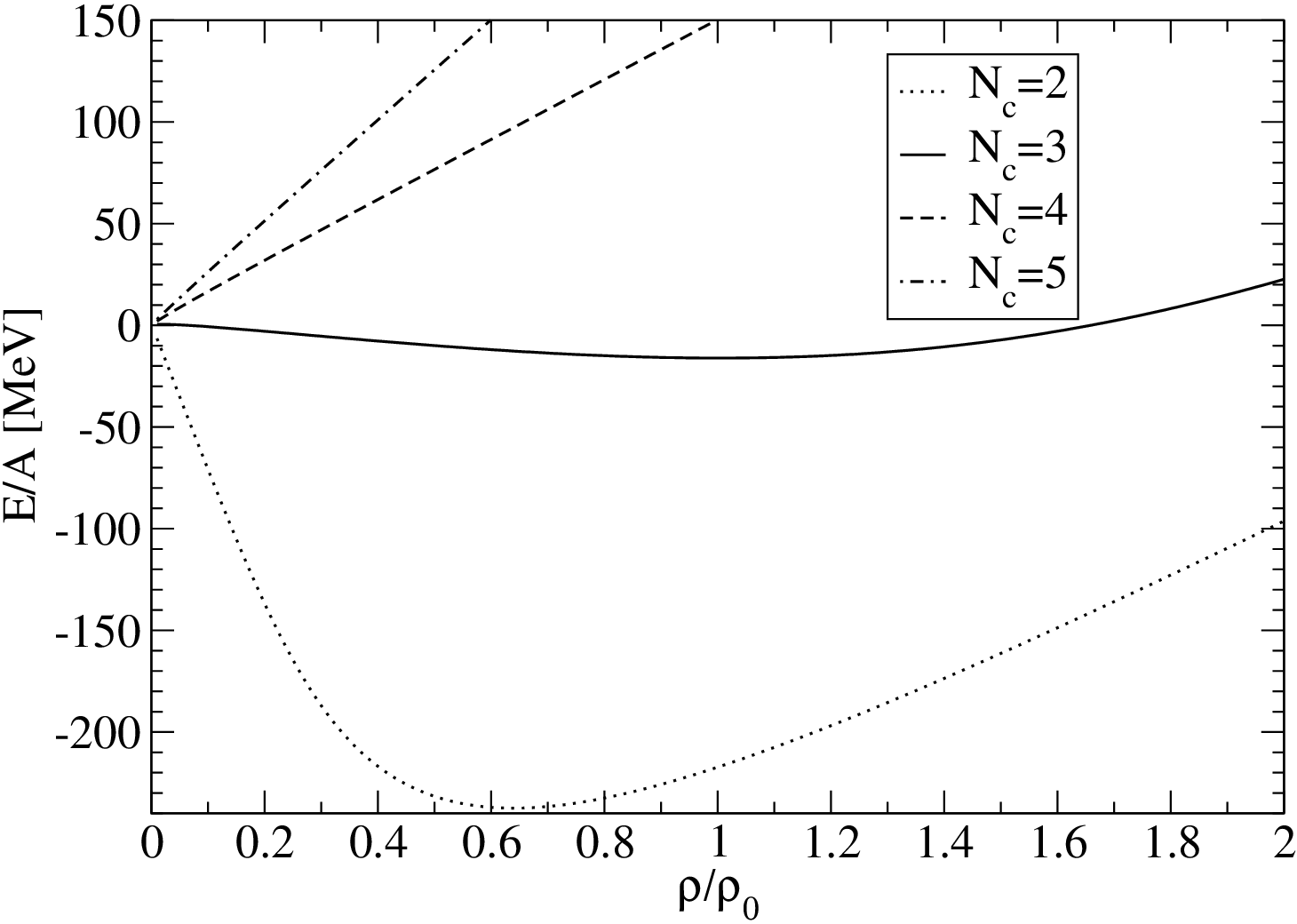} .\caption{Binding energy
per nucleon as a function of baryon density for the case in which the light
quark-antiquark $\sigma$ meson is replaced by a tetraquark state. In this case
it is clear that nuclear matter is unbound already for $N_{c}=4$.}%
\label{esuatetraq}%
\end{centering}
\end{figure}

We now repeat the study of the Walecka model by replacing the scalar state $%
\sigma $ with the tetraquark state $\chi .$ To this end we consider the
Walecka Lagrangian%
\begin{eqnarray}
\mathcal{L} &=&\bar{\psi}[\gamma ^{\mu }(i\partial _{\mu }-g_{\omega }\omega
_{\mu })-(m_{N}-g_{\chi }\chi )]\psi +\frac{1}{2}\partial ^{\mu }\chi
\partial _{\mu }\chi -\frac{1}{2}m_{\chi }^{2}\chi ^{2}  \notag \\
&-&\frac{1}{4}F^{\mu \nu }F_{\mu \nu }+\frac{1}{2}m_{\omega }^{2}\omega
^{\mu }\omega _{\mu }\text{ ,}
\end{eqnarray}%
with the following modified scaling properties: 
\begin{equation}
m_{\chi }\rightarrow m_{\chi }\frac{2N_{c}-2}{4}\text{ , }g_{\chi
}\rightarrow g_{\chi }\text{ .}  \label{mtqnc}
\end{equation}%
The numerical results are shown in Fig.~(\ref{esuatetraq}), where we plot
the binding energy per nucleon as a function of baryon density for different
values of $N_{c}$. As it is clear from the picture, the tetraquark scenario
leads to completely different results, compared with the previous scenario.
In particular, the nuclear matter binding energy decreases so fast with $%
N_{c}$ that already for $N_{c}=4$ no nuclear matter exists. This is a
consequence of the fact that, in this case, the medium-range attraction
between nucleons is mediated by the exchange of a tetraquark meson. When the
number of colors is increased, the tetraquark disappears, leading to a
strong weakening of the attraction between the nucleons. 

\subsection{The scenario with two scalar fields}

Even if the lightest scalar state $f_{0}(600)$ is mainly a tetraquark, one
still expects a chiral partner of the pion above 1 GeV which can be
interpreted as the resonance $f_{0}(1370)$ \cite{denis}. In the literature
scenarios with two scalar nonets have been investigated \cite{fariborz,tqmio}%
. Interestingly, in the detailed study of the nucleon-nucleon scattering
performed in Ref.~\cite{ml} two-scalar isoscalar states with masses $%
m_{\sigma_{1}}\sim400$-$600$ MeV and $m_{\sigma_{2}}\sim1200$ MeV have been
considered. In view of the previous discussion we interpret the light scalar
as predominantly tetraquark and the heavier one as predominantly quarkonium,
see also Ref.~\cite{heinz} in which the potentially important role of a
light tetraquark field at nonzero temperature has been outlined.

The natural question is if and how the present study is modified by the
presence of two scalar-isoscalar states. We thus consider an extended
version of the Walecka model with two scalar fields:%
\begin{align}
\mathcal{L} & =\bar{\psi}[\gamma^{\mu}(i\partial_{\mu}-g_{\omega}\omega_{\mu
})-(m_{N}-g_{\sigma}\sigma-g_{\chi}\chi)]\psi+\frac{1}{2}\partial^{\mu}%
\sigma\partial_{\mu}\sigma-\frac{1}{2}m_{\sigma}^{2}\sigma^{2}  \notag \\
& +\frac{1}{2}\partial^{\mu}\chi\partial_{\mu}\chi-\frac{1}{2}%
m_{\chi}^{2}\chi^{2}-\frac{1}{4}F^{\mu\nu}F_{\mu\nu}+\frac{1}{2}%
m_{\omega}^{2}\omega^{\mu}\omega_{\mu}\text{ .}  \label{walecka2}
\end{align}
It is interesting to notice that the general chiral model in Ref.~\cite%
{susanna} reduces to the Walecka form with two scalar fields in Eq.~(\ref%
{walecka2}) when nuclear matter properties are investigated and when the
glueball is neglected.

We use the following numerical values of the parameters: $%
g_{\chi}^{2}/4\pi=4.25$ with $m_{\chi}=470$ MeV, $g_{\sigma}^{2}/4\pi=17.61$
with $m_{\sigma}=1225$ MeV and $g_{\omega}^{2}/4\pi=14.28$ with $%
m_{\omega}=781$ MeV. The quantities $g_{\chi},$ $g_{\sigma}$ and $m_{\sigma}$
are directly taken from Ref. \cite{ml}. The parameters $m_{\chi}$ and $%
g_{\omega}$ have been fixed to obtain saturation at $\rho_{0}=0.16$ fm$^{-3}$
and an energy per nucleon $E/A=-16$ MeV. Interestingly, $m_{\chi}=470$ MeV
is only slightly changed with respect to the value of $452$ MeV reported in
Ref. \cite{ml}, while $g_{\omega}$ is $15\%$ smaller than the value Ref. 
\cite{ml}. (In view of the large uncertainty on the coupling $g_{\omega}$
this is not a large deviation, see also the discussion in Ref. \cite{ml2}).

We then perform the large $N_{c}$ study in accordance with Eqs.~(\ref%
{msigmanc}), (\ref{omega}), (\ref{gis}) and (\ref{mtqnc}). The numerical
results, depicted in Fig.~\ref{esuatetraq}, are quite similar to those
obtained in the previous scenario. Although the heavy quark-antiquark state
does not disappear with increasing $N_{c}$, its role alone is not sufficient
to bind nucleons together: once again, the increase of $N_{c}$ destroys
nuclear matter. It should be stressed that our results do not depend on fine
tuning: it is much more the nuclear matter for $N_{c}=3$ which follows from
the detailed balance.

\subsection{The scalar field as an effective treatment of TPE processes}

The one-pion-exchange (OPE) has not been considered up to now because it
does not contribute to the mean field approximation. Moreover, while the OPE
is surely important for the long-range attraction between two nucleons, it
is not enough to bind nuclei, see the review in Ref. \cite{baldo} and the
explicit study on the deuterium in Ref. \cite{deut} and refs. therein. A
middle-range attraction mediated by a scalar particle is a necessary
ingredient to describe nuclear matter.

However, one can go a step further and consider processes involving
two-pion-exchange \cite{weisenucl,arriola}, which generate the middle-range
attraction in the scalar-isoscalar channel. The description through a scalar
particle is then only an effective way to describe a process in which two
pions are exchanged between two nucleons.

In order to study how these contributions behave in the large $N_{c}$ limit
it is first necessary to review the OPE and its properties for $N_{c}\gg 3.$
The simplest OPE interaction term is given by: 
\begin{equation}
\mathcal{L}_{1,\pi }=-g_{\pi }^{(1)}\vec{\pi}\bar{\psi}\gamma ^{5}\vec{\tau}%
\psi \text{ ,}  \label{lp1}
\end{equation}%
Although $g_{\pi }^{(1)}\propto \sqrt{N_{c}}$ in the large $N_{c}$ limit,
the OPE-potential does not scale as $\left( g_{\pi }^{(1)}\right) ^{2}$ $%
\propto N_{c},$ but is proportional to $\left( g_{\pi }^{(1)}\right)
^{2}/M_{N}^{2}\propto N_{c}^{-1}$ and then it is suppressed when $N_{c}\gg 3$%
. This is due to the fact that, in virtue of the matrix $\gamma ^{5},$ an
additional factor $1/M_{N}$ is associated to the emission of one pion.
Notice that such a suppression is not present in the vector and scalar
channels: the corresponding potentials scale as $g_{\omega }^{2}\propto
N_{c} $ and $g_{scalar}^{2},$ whose scaling behavior depend on the
assignment for the scalar field, see the discussion above.

Another OPE\ interaction is however possible and involves the derivative of
the pion field:%
\begin{equation}
\mathcal{L}_{2,\pi }=-g_{\pi }^{(2)}\left( \partial _{\mu }\vec{\pi}\right) 
\bar{\psi}\gamma ^{\mu }\gamma ^{5}\vec{\tau}\psi \text{ .}  \label{lp2}
\end{equation}%
Unlike the previous case, the corresponding OPE-potential is simply
proportional to $\left( g_{\pi }^{(2)}\right) ^{2}$. It is then crucial to
understand how $g_{\pi }^{(2)}$ scales in the large $N_{c}$ limit. If $%
g_{\pi }^{(2)}$ scales, as naively expected, as $\sqrt{N_{c}},$ then such a
contribution to the OPE-potential survives for $N_{c}\gg 3$.

From the perspective of soft-pion emission, the Lagrangian $\mathcal{L}%
_{1,\pi }$ in Eq. (\ref{lp1}) can be equivalently replaced by a Lagrangian
of the form of Eq. (\ref{lp2}) in the following way:%
\begin{equation}
\mathcal{L}_{1,\pi }\rightarrow -\frac{g_{\pi }^{(1)}}{2M_{N}}\left(
\partial _{\mu }\vec{\pi}\right) \bar{\psi}\gamma ^{\mu }\gamma ^{5}\vec{\tau%
}\psi \text{ . }
\end{equation}%
Clearly, the large $N_{c}$ behavior is not changed, being still $\left(
g_{\pi }^{(1)}/M_{N}\right) ^{2}\propto N_{c}^{-1}.$

In linear sigma models without (axial-)vector mesons only the pion-nucleon
interaction of the form given in Eq. (\ref{lp1}) is present. Moreover,
neglecting the small contribution from the nonzero current quark masses, the
nucleon mass takes the form $M_{N}=g_{\pi }^{(1)}f_{\pi },$ thus one
obtains: 
\begin{equation}
\mathcal{L}_{1,\pi }\rightarrow -\frac{1}{2f_{\pi }}\left( \partial _{\mu }%
\vec{\pi}\right) \bar{\psi}\gamma ^{\mu }\gamma ^{5}\vec{\tau}\psi \text{ ,}
\end{equation}%
where the scaling is not changed since $f_{\pi }\propto \sqrt{N_{c}}.$

In general, however, the full pion-nucleon interaction in the soft-pion
limit takes the form: 
\begin{equation}
\mathcal{L}_{\pi ,full}=\mathcal{L}_{1,\pi }+\mathcal{L}_{2,\pi }\rightarrow
-\left( \frac{g_{\pi }^{(1)}}{2M_{N}}+g_{\pi }^{(2)}\right) \left( \partial
_{\mu }\vec{\pi}\right) \bar{\psi}\gamma ^{\mu }\gamma ^{5}\vec{\tau}\psi 
\text{ . }
\end{equation}%
In chiral perturbation theory \cite{weisenucl}, which is also defined in the
soft-pion limit, the following pion-nucleon interaction term is present: 
\begin{equation}
\mathcal{L}_{\pi }^{chPT}=-\frac{g_{A}}{2f_{\pi }}\left( \partial _{\mu }%
\vec{\pi}\right) \bar{\psi}\gamma ^{\mu }\gamma ^{5}\vec{\tau}\psi \text{ }
\end{equation}%
where $g_{A}$ is the axial-coupling of the nucleon. By comparison we find 
\begin{equation}
g_{A}=2f_{\pi }\left( \frac{g_{\pi }^{(1)}}{2M_{N}}+g_{\pi }^{(2)}\right)
=1+2f_{\pi }g_{\pi }^{(2)}\text{.}
\end{equation}%
In the already mentioned sigma models without vector mesons the constant $%
g_{\pi }^{(2)}=0,$ and one has therefore $g_{A}=1\propto N_{c}^{0}.$ The
situation changes, however, when vector mesons are included: in those models 
$g_{\pi }^{(2)}$ does not vanish and scales as $\sqrt{N_{c}}.$ This, in
turn, implies that $g_{A}\propto N_{c}.$ The reason why a term of the kind $%
g_{\pi }^{(2)}\left( \partial _{\mu }\vec{\pi}\right) \bar{\psi}\gamma ^{\mu
}\gamma ^{5}\vec{\tau}\psi $ with $g_{\pi }^{(2)}\propto \sqrt{N_{c}}$
emerges is rather subtle: it follows from the so-called $a_{1}$-$\pi $
mixing, see Ref. \cite{susanna} for details. On a numerical level, in htese
generalized sigma models the following large $N_{c}$ scaling for the
axial-coupling constant $g_{A}$ is obtained:%
\begin{equation}
g_{A}(N_{c})=1+0.267\frac{N_{c}}{3}=1+\left( g_{A}^{\exp }-1\right) \frac{%
N_{c}}{3}  \label{ganc}
\end{equation}%
where $g_{A}^{\exp }=1.267\pm 0.004$. There is therefore a factor $1$ which
is large $N_{c}$ independent and a factor $\left( g_{A}^{\exp }-1\right)
N_{c}/3=0.267N_{c}/3,$ which is subdominant for $N_{c}=3$ but which
dominates for $N_{c}$ large enough\footnote{%
When the so called mirror assignment for the nucleon and its chiral partner
is considered in the framework of the linear sigma models \cite%
{susanna,detar}, Eq. (\ref{ganc}) changes as $g_{A}(N_{c})=a+bN_{c}/3,$
where $a$ is not necessarily unity. However, the large $N_{c}$ behavior is
unaffected and, using the numerical results of Ref. \cite{susanna}, the
result $g_{A}(N_{c})=0.93+0.33N_{c}/3$ is obtained, which is only slightly
changed w.r.t. Eq. (\ref{ganc}). All the conclusions presented in this
section hold therefore also in the mirror assignment.}. Thus, the deviation
of the axial coupling constant form unity becomes dominant in the large $%
N_{c}$ limit. This discussion shows that a OPE-term does not disappear in
the large $N_{c}$ limit, although its strength with respect to the $\omega $
meson repulsion is smaller than in the vacuum.

We now turn to the TPE case, which can mimic the middle-range scalar
attraction. When considering two pions as intermediate state, the naive
scaling of the corresponding TPE potential is proportional to $\left(
g_{A}/2f_{\pi }\right) ^{2}\propto N_{c}^{2}.$ This result is however not
correct: the $N_{c}^{2}$ contribution from the box diagrams cancel with the $%
N_{c}^{2}$ contribution from the crossed-box diagram, see Ref.\cite{cohen}
for details. The resulting TPE potential is then proportional to $\left(
g_{A}/2f_{\pi }\right) ^{2}/N_{c}\propto N_{c}.$

Now, when substituting the TPE interactions with an effective scalar field $%
\sigma _{TPE}$ (where the effective nature of this field, in comparison to
the quarkonium and tetraquark cases, should be clear) we have the following
Walecka-type Lagrangian 
\begin{eqnarray}
\mathcal{L} &=&\bar{\psi}[\gamma ^{\mu }(i\partial _{\mu }-g_{\omega }\omega
_{\mu })-(m_{N}-g_{\sigma _{TPE}}\sigma _{TPE})]\psi +\frac{1}{2}\partial
^{\mu }\sigma _{TPE}\partial _{\mu }\sigma _{TPE}  \notag \\
&-&\frac{1}{2}m_{\sigma _{TPE}}^{2}\sigma _{TPE}^{2}\text{ ,}
\end{eqnarray}%
with 
\begin{equation}
g_{\sigma _{TPE}}^{2}(N_{c})=\left[ \frac{g_{A}(N_{c})}{g_{A}(N_{c}=3)}%
\right] ^{2}\frac{3g_{\sigma _{TPE}}^{2}(N_{c}=3)}{N_{c}}  \label{spp}
\end{equation}%
and 
\begin{equation}
m_{\sigma _{TPE}}^{2}\sim 4m_{\pi }^{2}\propto N_{c}^{0}\text{ .}
\label{mspp}
\end{equation}%
Notice that $g_{\sigma _{TPE}}^{2}(N_{c})$ is chosen in such a way that, as
usual, for $N_{c}=3$ the value $g_{\sigma _{TPE}}^{2}(N_{c}=3)/4\pi =2.418$
necessary for the experimentally observed saturation is realized.

In order to test if nuclear matter exist for large $N_{c}$ in this scenario
we repeat our study by using the new scaling in Eqs. (\ref{spp}) and (\ref%
{mspp}). Although $g_{\sigma _{TPE}}$ scales with $N_{c}$ (as in the
quarkonium assignment), no saturation is obtained in the large $N_{c}$
limit. The reason for this result is that the ratio $g_{\sigma
_{TPE}}/g_{\omega }$ decreases as soon as $N_{c}=3$ is left and then
approaches a constant for $N_{c}\rightarrow \infty $, which is however
smaller than the value $\left( g_{\sigma _{TPE}}/g_{\omega }\right)
_{N_{c}=3}.$ For instance, we can obtain a stable nuclear matter for $N_{c}=4
$ (but not for larger values of $N_{c}$) only if we would --artificially--
use a large value $g_{A}^{\exp }\gtrsim 5$. This is, however, not the case
in our world, where $g_{A}^{\exp }=1.267.$

We thus conclude that, also when the scalar attraction is mediated by TPE
processes in the scalar channel, nuclear matter does not bind for large $%
N_{c}.$ This result holds true even when $g_{A}$ scales as $N_{c},$ provided
that the axial coupling constant measured in the real world for $N_{c}=3$ is
reproduced.

\subsection{Further scenarios}

\begin{itemize}
\item Dilaton/Glueball: In this work we did not consider the glueball field
as a possible intermediate boson for the nucleon-nucleon interaction.
Although potentially important in dilatation invariant models and for the
scalar phenomenology \cite{varieglue}, the mass of the glueball is about $%
1.5 $ GeV \cite{lattglue} and is too high to affect nuclear matter binding.
There are however models in which the lightest scalar resonance $f_{0}(600)$
is interpreted as a glueball state, e.g.~\cite{vento} and refs.~therein.
Although we consider this assignment unfavored due to the too low glueball
mass in comparison with the lattice value, for completeness we perform a
study of this scenario. The corresponding Lagrangian reads 
\begin{eqnarray}
\mathcal{L}&=&\bar{\psi}[\gamma^{\mu}(i\partial_{\mu}-g_{\omega}\omega_{\mu
})-(m_{N}-g_{G}G)]\psi+\frac{1}{2}\partial^{\mu}G\partial_{\mu}G-\frac{1}{2}%
m_{G}^{2}G^{2}  \notag \\
&-&\frac{1}{4}F^{\mu\nu}F_{\mu\nu}+\frac{1}{2}m_{\omega}^{2}\omega^{\mu}%
\omega_{\mu}\text{ .}
\end{eqnarray}
The leading order large $N_{c}$ scaling relations are given by 
\begin{equation}
m_{G}\longrightarrow m_{G}\text{ , }g_{G}\rightarrow g_{G}\text{ ,}
\end{equation}
i.e.~they are both large $N_{c}$ invariant. When repeating the study in this
case no bound state in the large $N_{c}$ scenario exists.

\item In many works on light hadron states it is found that light scalars
are `dynamically generated' \cite{pelaeznc}, see also the discussion about
dynamically generated and reconstructed states in Ref.~\cite{dynrec}. In
particular, the light resonance $f_{0}(600)$ does not correspond to any of
the previously analyzed cases (quark-antiquark, tetraquark or glueball) but
is a pion-pion `molecular' bound state. Notice that the here analyzed
scenario is also different from the TPE case studied in Sec. 2.7, where two
pions were simultaneously exchanged between two nucleons, without
interacting with each other. Although the emission of the two pions is
necessary in the present case, the crucial point here is their further
interaction to generate a new resonance. This new resonance, contrary to the
simple TPE processes, definitely disappears in the large $N_{c}$ limit. The
reason is that the attraction in the $\pi \pi $ channel is mediated by meson
exchange (such as the exchange of a $\rho $ meson), whose corresponding
scattering amplitudes scale as $1/N_{c}.$ For $N_{c}$ large enough the
interaction strength fades out and the light scalar resonance ceases to
exist. Thus, even its effect for nuclear matter does not take place: no
binding at large $N_{c}$ takes place in scenarios in which $f_{0}(600)$ is
dynamically generated.

\item The previous conclusion holds also when the light resonance $%
f_{0}(600) $ emerges as a low-energy companion pole \cite{pennington}. In
fact, these new poles disappear for $N_{c}\gg 3$.
\end{itemize}

\section{Conclusions}

In this work we studied the formation of nuclear matter for $N_{c}\gg3.$ We
conclude that the present phenomenological information about scalar mesons
implies that no nuclear matter exists for large $N_{c}$. In fact, the only
case in which nuclear matter does not disappear by increasing $N_{c}$ is the
naive quarkonium assignment for the lightest scalar resonance. This scenario
is criticized by many recent and less recent studies of low-energy hadron
phenomenology, which agree that the light scalar states below 1 GeV are not
predominantly quarkonium states. Moreover, even in the quarkonium picture
one should at least include the effects of the pion-pion dressing, which is
expected to be large in view of the broad nature of the resonance. This
property is enough to `unbind' nuclear matter for large $N_{c}.$

The non existence of nuclear matter for large $N_{c}$ has been explicitly
shown in alternative scenarios for the light scalar states. We first
concentrated on the tetraquark interpretation, in which a peculiar large $%
N_{c}$ limit has been discussed. We then studied the cases in which the
nucleon-nucleon interaction in the scalar channel is dominated by: (i) two
scalar fields, (ii) TPE processes, (iii) a glueball state, and (iv) a
pion-pion molecular state. The common feature of all these assignments is
that nuclear matter does not bind for large $N_{c}.$ Numerically, the value $%
N_{c}=4$ is already enough to render nuclear matter unstable.

The results of this work have been derived by using Walecka-type models. We
have limited the study to nuclear matter density and small temperatures,
where the Walecka model represents a well-defined and useful theoretical
tool. Moreover, the main goal of the present work is not a precise numerical
study of nuclear matter properties, but simply to assess its existence for $%
N_{c}\gg 3$: for this reason a simple and schematic model as the Walecka one
fulfills the desired requirements. However, it is surely an interesting task
for the future to repeat the present study going beyond the mean field
approximations used here.

Obviously we do live in a world in which nuclear matter exists. The laws of
Nature must allow for nuclear matter, so that life as we know it might
evolve (anthropic principle). The subtle point is not (only) the existence
of nuclear matter, but the fact that the binding energy per nucleon $%
E_{B}\simeq 16$ MeV is much smaller than the natural scale of the system, $%
\Lambda _{QCD}\simeq 200$ MeV. Our outcome that nuclear matter exists only
for $N_{c}\lesssim 3,$ and is thus a peculiar property of our world, is in
agreement with the phenomenological realized smallness of the ratio $%
E_{B}/\Lambda _{QCD}\simeq 0.1$. In fact, in the tetraquark (or molecular)
scenario, the decreasing of $N_{c}$ favors the formation of nuclear matter.
By decreasing $N_{c}$ from $3$ to $2$ an increase of the binding energy is
obtained. In the framework of nuclear matter, $N_{c}=3$ is not large at all.
It should be stressed that all this is not true when the lightest scalar
state is a quarkonium state, for which the relation $E_{B}/\Lambda
_{QCD}\sim N_{c}$ holds. In this (unfavored) scenario the smallness of the
binding energy could not be understood. Note that, while it is clearly not
possible to investigate experimentally the (non)binding of nuclear matter
for $N_{c}>3,$ this can be the subject of computer simulations of QCD, in
which the number of colors is a parameter which can be easily changed.

Many studies have already shown that the conditions for the existence of
complex life represent a small volume in the space of the free parameters
(coupling constants and masses) of the Standard Model (e.g.~\cite{jenkins}
and refs.~therein). The present study shows that these conditions are
restricted also in the direction of $N_{c}.$ The change of $N_{c}$ can be
regarded as a change of the group structure of the standard model. The fact
that the group of the strong interaction in our Universe is $SU(N_{c})$ with 
$N_{c}\lesssim3$ should not be a surprise.

\bigskip

\textbf{Acknowledgement: }The authors thank Daniel Fernandez-Freile,
G.~Pagliara, A.~Heinz, G.~Torrieri and S.~Lottini for useful discussions.
L.~B.~is supported by the Hessen Initiative for Excellence (LOEWE) through
the Helmholtz International Center for FAIR (HIC for FAIR).


\bigskip

\begin{thebibliography}{99}
\bibitem{thooft} G.~'t Hooft, 
Nucl.\ Phys.\ B \textbf{72} (1974) 461. 

\bibitem{witten} E.~Witten, 
Nucl.\ Phys.\ B \textbf{160} (1979) 57. 

\bibitem{reviewnc} S.~R.~Coleman, ``1/N,'' Published in Erice Subnuclear
1979:0011. 
R.~F.~Lebed, 
Czech.\ J.\ Phys.\ \textbf{49} (1999) 1273 [arXiv:nucl-th/9810080]. 
E.~E.~Jenkins, 
Ann.\ Rev.\ Nucl.\ Part.\ Sci.\ \textbf{48} (1998) 81
[arXiv:hep-ph/9803349]. 


\bibitem{quarkyonic} L.~McLerran and R.~D.~Pisarski, 
Nucl.\ Phys.\ A \textbf{796} (2007) 83 [arXiv:0706.2191 [hep-ph]]. 

\bibitem{sasaki} L.~McLerran, K.~Redlich and C.~Sasaki, 
arXiv:0812.3585 [hep-ph]. 
C.~Sasaki and I.~Mishustin, 
Phys.\ Rev.\ C \textbf{82} (2010) 035204 [arXiv:1005.4811 [hep-ph]]. 

\bibitem{torrieri} G.~Torrieri and I.~Mishustin, 
Phys.\ Rev.\ C \textbf{82} (2010) 055202 [arXiv:1006.2471 [nucl-th]]. 
G.~Torrieri and I.~Mishustin, 
arXiv:1101.0149 [nucl-th]. 

\bibitem{dicoto} Y.~Hidaka, T.~Kojo, L.~McLerran and R.~D.~Pisarski, 
arXiv:1004.2261 [hep-ph]. 

\bibitem{kojo} T.~Kojo, Y.~Hidaka, L.~McLerran and R.~D.~Pisarski, 
Nucl.\ Phys.\ A \textbf{843} (2010) 37 [arXiv:0912.3800 [hep-ph]]. 

\bibitem{walecka} J.~D.~Walecka, 
Annals Phys.\ \textbf{83}, 491 (1974). 
B.~D.~Serot and J.~D.~Walecka, 
Adv.\ Nucl.\ Phys.\ \textbf{16}, 1 (1986). 
B.~D.~Serot and J.~D.~Walecka, 
Int.\ J.\ Mod.\ Phys.\ E \textbf{6}, 515 (1997) [arXiv:nucl-th/9701058]. 

\bibitem{amslerrev} C.~Amsler and N.~A.~Tornqvist, 
Phys.\ Rept.\ \textbf{389}, 61 (2004). 
E.~Klempt and A.~Zaitsev, 
Phys.\ Rept.\ \textbf{454} (2007) 1 [arXiv:0708.4016 [hep-ph]]. 

\bibitem{dynrec} F.~Giacosa, 
Phys.\ Rev.\ D \textbf{80} (2009) 074028 [arXiv:0903.4481 [hep-ph]]. 

\bibitem{pelaeznc} J.~R.~Pelaez, 
Phys.\ Rev.\ Lett.\ \textbf{92} (2004) 102001. 
Mod.\ Phys.\ Lett.\ A \textbf{19} (2004) 2879. 
M.~Uehara, 
arXiv:hep-ph/0401037. 
J.~A.~Oller and E.~Oset, 
Nucl.\ Phys.\ A \textbf{620} (1997) 438 [Erratum-ibid.\ A \textbf{652}
(1999) 407] [arXiv:hep-ph/9702314]. 
F.~Giacosa and G.~Pagliara, 
Nucl.\ Phys.\ A \textbf{833} (2010) 138 [arXiv:0905.3706 [hep-ph]]. 

\bibitem{pennington} 
M.~Boglione and M.~R.~Pennington, 
Phys.\ Rev.\ D \textbf{65} (2002) 114010 [arXiv:hep-ph/0203149]. 
E.~van Beveren, T.~A.~Rijken, K.~Metzger, C.~Dullemond, G.~Rupp and
J.~E.~Ribeiro, 
Z.\ Phys.\ C \textbf{30} (1986) 615 [arXiv:0710.4067 [hep-ph]].

\bibitem{denis} D.~Parganlija, F.~Giacosa and D.~H.~Rischke, 
Phys.\ Rev.\ D \textbf{82} (2010) 054024 [arXiv:1003.4934 [hep-ph]]. 

\bibitem{jaffeorig} R.~L.~Jaffe, 
Phys.\ Rev.\ D \textbf{15} (1977) 267. 
R.~L.~Jaffe, 
Phys.\ Rev.\ D \textbf{15} (1977) 281. 

\bibitem{exotica} R.~L.~Jaffe, 
Phys.\ Rept.\ \textbf{409} (2005) 1 [Nucl.\ Phys.\ Proc.\ Suppl.\ \textbf{142%
} (2005) 343] [arXiv:hep-ph/0409065]. 


\bibitem{varietq} 
L.~Maiani, F.~Piccinini, A.~D.~Polosa and V.~Riquer, 
Phys.\ Rev.\ Lett.\ \textbf{93} (2004) 212002 [arXiv:hep-ph/0407017].
F.~Giacosa, 
Phys.\ Rev.\ D \textbf{74} (2006) 014028 [arXiv:hep-ph/0605191]. 

\bibitem{fariborz} 
A.~H.~Fariborz, R.~Jora and J.~Schechter, 
Phys.\ Rev.\ D \textbf{72} (2005) 034001 [arXiv:hep-ph/0506170]. 
A.~H.~Fariborz, 
Int.\ J.\ Mod.\ Phys.\ A \textbf{19} (2004) 2095. [arXiv:hep-ph/0302133]. 
M.~Napsuciale and S.~Rodriguez, 
Phys.\ Rev.\ D \textbf{70} (2004) 094043. 

\bibitem{tqmio} F.~Giacosa, 
Phys.\ Rev.\ D \textbf{75} (2007) 054007 [arXiv:hep-ph/0611388]. 

\bibitem{ml} R.~Machleidt, 
Phys.\ Rev.\ C \textbf{63} (2001) 024001 [arXiv:nucl-th/0006014]. 


\bibitem{weisenucl} N.~Kaiser, R.~Brockmann and W.~Weise, 
Nucl.\ Phys.\ A \textbf{625} (1997) 758 [arXiv:nucl-th/9706045]. 
N.~Kaiser, S.~Gerstendorfer and W.~Weise, 
Nucl.\ Phys.\ A \textbf{637} (1998) 395 [arXiv:nucl-th/9802071]. 
L.~Girlanda, A.~Rusetsky and W.~Weise, 
Annals Phys.\ \textbf{312} (2004) 92 [arXiv:hep-ph/0311128]. 

\bibitem{manohar} D.~B.~Kaplan and A.~V.~Manohar, 
Phys.\ Rev.\ C \textbf{56} (1997) 76 [arXiv:nucl-th/9612021]. 
M.~M.~Kaskulov and H.~Clement, 
Phys.\ Rev.\ C \textbf{70} (2004) 014002 [arXiv:nucl-th/0401061]. 
S.~R.~Beane, 
arXiv:hep-ph/0204107. 


\bibitem{arriola} A.~Calle Cordon and E.~Ruiz Arriola, 
Phys.\ Rev.\ C \textbf{80} (2009) 014002 [arXiv:0904.0421 [nucl-th]]. 
A.~Calle Cordon and E.~Ruiz Arriola, 
Phys.\ Rev.\ C \textbf{81} (2010) 044002 [arXiv:0905.4933 [nucl-th]]. 

\bibitem{lee} B.W.\ Lee, \textquotedblleft Chiral
Dynamics\textquotedblright, Gordon and Breach, New York, 1972. 

\bibitem{geffen} S.~Gasiorowicz and D.~A.~Geffen, 
Rev.\ Mod.\ Phys.\ \textbf{41} (1969) 531. 


\bibitem{njl} Y.~Nambu and G.~Jona-Lasinio, 
Phys.\ Rev.\ \textbf{122} (1961) 345. 
For reviews: T.~Hatsuda and T.~Kunihiro, 
Phys.\ Rept.\ \textbf{247}, 221 (1994) [arXiv:hep-ph/9401310]. 
S.~P.~Klevansky, 
Rev.\ Mod.\ Phys.\ \textbf{64} (1992) 649. 

\bibitem{lupo} F.~Giacosa and G.~Pagliara, 
Phys.\ Rev.\ C \textbf{76} (2007) 065204 [arXiv:0707.3594 [hep-ph]]. 

\bibitem{variedq} A.~De Rujula, H.~Georgi and S.~L.~Glashow, 
Phys.\ Rev.\ D \textbf{12} (1975) 147. 
T.~A.~DeGrand, R.~L.~Jaffe, K.~Johnson and J.~E.~Kiskis, 
Phys.\ Rev.\ D \textbf{12} (1975) 2060. 
G.~'t Hooft, 
Phys.\ Rev.\ D \textbf{14}, 3432 (1976) [Erratum-ibid.\ D \textbf{18}, 2199
(1978)]. 
E.~V.~Shuryak, 
Nucl.\ Phys.\ B \textbf{203} (1982) 93. 
T.~Schafer and E.~V.~Shuryak, 
Rev.\ Mod.\ Phys.\ \textbf{70} (1998) 323 [arXiv:hep-ph/9610451]. 
E.~Shuryak and I.~Zahed, 
Phys.\ Lett.\ B \textbf{589} (2004) 21 [arXiv:hep-ph/0310270]. 
U.~Vogl and W.~Weise, 
Prog.\ Part.\ Nucl.\ Phys.\ \textbf{27} (1991) 195. 
P.~Maris and C.~D.~Roberts, 
Int.\ J.\ Mod.\ Phys.\ E \textbf{12} (2003) 297 [arXiv:nucl-th/0301049]. 

\bibitem{liu} C.~Liu, 
Eur.\ Phys.\ J.\ C \textbf{53} (2008) 413 [arXiv:0710.4185 [hep-ph]]. 

\bibitem{heinz} A.~Heinz, S.~Struber, F.~Giacosa and D.~H.~Rischke, 
Phys.\ Rev.\ D \textbf{79} (2009) 037502 [arXiv:0805.1134 [hep-ph]]. 

\bibitem{susanna} S.~Gallas, F.~Giacosa and D.~H.~Rischke, 
Phys.\ Rev.\ D \textbf{82} (2010) 014004 [arXiv:0907.5084 [hep-ph]]. 

\bibitem{ml2} R.~Machleidt, K.~Holinde and C.~Elster, 
Phys.\ Rept.\ \textbf{149} (1987) 1. 



\bibitem{baldo} M.~Baldo and G.~F.~Burgio, 
arXiv:1102.1364 [nucl-th]. 

\bibitem{deut} Y.~B.~Ding \textit{et al.}, 
J.\ Phys.\ G \textbf{30} (2004) 841 [arXiv:hep-ph/0402109]. 

\bibitem{detar} C.~DeTar and T.~Kunihiro, 
Phys.\ Rev.\ D \textbf{39} (1989) 2805. 


\bibitem{cohen} M.~K.~Banerjee, T.~D.~Cohen and B.~A.~Gelman, 
Phys.\ Rev.\ C \textbf{65} (2002) 034011 [arXiv:hep-ph/0109274]. 

\bibitem{varieglue} 
C.~Amsler and F.~E.~Close, 
Phys.\ Rev.\ D \textbf{53} (1996) 295 [arXiv:hep-ph/9507326]. 
W.~J.~Lee and D.~Weingarten, 

Phys.\ Rev.\ D \textbf{61}, 014015 (2000). [arXiv:hep-lat/9910008]; 
F.~E.~Close and A.~Kirk, 
Eur.\ Phys.\ J.\ C \textbf{21}, 531 (2001). [arXiv:hep-ph/0103173]. 
F.~Giacosa, T.~Gutsche, V.~E.~Lyubovitskij and A.~Faessler, 
Phys.\ Rev.\ D \textbf{72}, 094006 (2005). [arXiv:hep-ph/0509247]. 
F.~Giacosa, T.~Gutsche, V.~E.~Lyubovitskij and A.~Faessler, 
Phys.\ Lett.\ B \textbf{622}, 277 (2005) [arXiv:hep-ph/0504033]. 
F.~Giacosa, T.~Gutsche and A.~Faessler, 
Phys. Rev. C \textbf{71}, 025202 (2005) [arXiv:hep-ph/0408085]. 
H.~Y.~Cheng, C.~K.~Chua and K.~F.~Liu, 
Phys.\ Rev.\ D \textbf{74} (2006) 094005 [arXiv:hep-ph/0607206]. 
L.~Bonanno and A.~Drago, 
Phys.\ Rev.\ C \textbf{79}, 045801 (2009) [arXiv:0805.4188 [nucl-th]]. 


\bibitem{lattglue} Y.~Chen \textit{et al.}, 
Phys.\ Rev.\ D \textbf{73} (2006) 014516. [arXiv:hep-lat/0510074]. 

\bibitem{vento} V.~Mathieu, N.~Kochelev and V.~Vento, 
Int.\ J.\ Mod.\ Phys.\ E \textbf{18} (2009) 1 [arXiv:0810.4453 [hep-ph]]. 

\bibitem{jenkins} R.~L.~Jaffe, A.~Jenkins and I.~Kimchi, 
Phys.\ Rev.\ D \textbf{79} (2009) 065014 [arXiv:0809.1647 [hep-ph]]. 
R.~Kallosh and A.~D.~Linde, 
Phys.\ Rev.\ D \textbf{67} (2003) 023510 [arXiv:hep-th/0208157]. 
V.~Agrawal, S.~M.~Barr, J.~F.~Donoghue and D.~Seckel, 
Phys.\ Rev.\ D \textbf{57} (1998) 5480 [arXiv:hep-ph/9707380]. 

\end{thebibliography}
\end{document}